\documentclass[lettersize,journal]{IEEEtran}
\usepackage{amsmath,amsfonts}
\usepackage{algorithmic}
\usepackage{algorithm}
\usepackage{array}
\usepackage[caption=false,font=normalsize,labelfont=sf,textfont=sf]{subfig}
\usepackage{textcomp}
\usepackage{stfloats}
\usepackage{url}
\usepackage{fancyhdr}
\usepackage{verbatim}
\usepackage{graphicx}
\usepackage{cite}
\usepackage{etoolbox}
\makeatletter
\patchcmd{\@makecaption}
  {\scshape}
  {}
  {}
  {}
\makeatother
\fancypagestyle{FirstPage}{
\lfoot{978-1-6654-8453-4/22/\$31.00 ©2022 IEEE} 
}
\begin{document}
\title{NFDLM: A Lightweight Network Flow based Deep Learning Model for DDoS Attack Detection in IoT Domains}
\author{ Kumar Saurabh$^1$, Tanuj Kumar$^1$, Uphar Singh$^1$, O.P. Vyas$^1$, Rahamatullah Khondoker$^2$ \\$^1$Indian Institute of Information Technology, Allahabad, India\\$^2$Department of Business Informatics, THM University of Applied Sciences, Friedberg, Germany\\ pwc2017001@iiita.ac.in, mit2020075@iiita.ac.in, pse2017003@iiita.ac.in, opvyas@iiita.ac.in,  rahamatullah.khondoker@mnd.thm.de}
\IEEEpubid{\makebox[\columnwidth]{978-1-6654-8453-4/22/\$31.00 ©2022 IEEE\hfill} \hspace{\columnsep}\makebox[\columnwidth]{ }}
\maketitle
\vspace{-2pt}
\begin{abstract}
In the recent years, Distributed Denial of Service (DDoS) attacks on Internet of Things (IoT) devices have become one of the prime concerns to Internet users around the world. One of the sources of the attacks on IoT ecosystems are botnets. Intruders force IoT devices to become unavailable for its legitimate users by sending large number of messages within a short interval. This study proposes NFDLM, a lightweight and optimised Artificial Neural Network (ANN) based Distributed Denial of Services (DDoS) attack detection framework with mutual correlation as feature selection method which produces a superior result when compared with Long Short Term Memory (LSTM) and simple ANN. Overall, the detection performance achieves approximately 99\% accuracy for the detection of attacks from botnets. In this work, we have designed and compared four different models where two are based on ANN and the other two are based on LSTM to detect the attack types of DDoS. 
\end{abstract}

\thispagestyle{FirstPage}
\vspace{-0.5cm}
\section{Introduction}
\IEEEPARstart The Internet of Things (IoT) is a network of interconnected smart gadgets that uses various services to connect them into a single network~\cite{xiao2018iot}. It enables smart devices to collect sensitive data, perform critical operations, interact and communicate with each other. Various malicious activities such as unlawful data access, stolen credentials, impersonation, data tampering, infiltration have been seen increasing at alarming rate in recent years~\cite{hafeez2020iot},~\cite{tian2019distributed}. Existing security and preventative approaches are no longer adequate to give total security against complex threats and viruses that are constantly changing, making detection more difficult and complicated~\cite{ahmad2021machine}. Malicious operations that aim to interrupt a service, commonly known as Denial of Service (DoS), occurs when a website is rendered unavailable to genuine users~\cite{ravi2020learning,kolias2017ddos}. DDoS are carried out by a group of infected devices known as “Bots", which are automated programs that attack on a distant system (generally a server) ~\cite{pimenta2017cybersecurity}. A DoS attack happens when a hacker disrupts the services of a network connected host in order to prevent legitimate users from accessing a server or network resource. When multiple sources attack the compromised server, the assault is known as “Distributed Denial of Service" (DDOS) attack~\cite{fortinet_report}. The goal of such attacks is to cause disruption having the potential to be life-threatening. Major contributors of attacks in IoT are DoS and DDoS. These assaults may be volumetric or protocol based~\cite{zargar2013survey}. A large number of people are affected by volumetric assaults. These assaults produce a huge amount of network traffic by either forcing the victim to carry out the attacker's commands or cause the system to crash, rendering the service unavailable. SYN flood assault, ICMP flood assault, and UDP flood assault~\cite{gupta2020detecting} are all common volumetric DDoS assaults. Protocol based attacks make use of Transport Control Protocol (TCP) / Internet Protocol (IP) or User Datagram Protocol (UDP) / IP mechanics to exhaust the CPU and memory resources, rendering the targeted system unable to respond to queries, Fig 1. shows the DDoS attack scenario.
\begin{figure}[h!]
\centering
\includegraphics[width=3.5in]{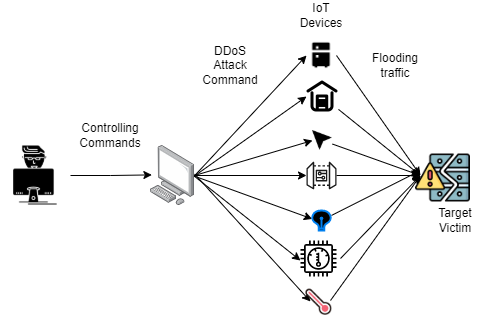}
\caption{DDoS attack scenario diagram}
\label{fig_1}
\end{figure}

This paper discusses four novel deep learning based architectures for DDoS attack detection in IoT networks. These models enable classification with less resources, faster training time and higher accuracy. The following points outline the novelty of our work:
\begin{enumerate}
  \item This paper presents an overview of the widely used Deep Learning (DL) technologies and potential DDoS attacks in IoT networks.
  \item Four different feature selection-based DL techniques are proposed with Correlation and Mutual Information based feature selection techniques to ensure higher attack detection accuracy for DDoS attack detection in IoT networks.
  \item This paper gives a lightweight and highly accurate model for DDoS attack detection in IoT networks. 
\end{enumerate}The remainder of this paper is organized as follows:
“Related Work” investigates the widely used Machine Learning (ML) and Deep Lerning (DL) techniques and the open issues for IoT cybersecurity. “Overview of Deep Learning and Background Methodologies" discusses about the LSTM, ANN and other background methodologies. “Proposed Methodology and Experimental Setup" section describes the proposed IDS design and the steps involved from design and data preprocessing to model training. “Results and Discussion" section deals with the brief discussion on feature selection and experimental findings. Finally, the “Conclusion and Future Research" section discusses the summary of our work and future directions.
\vspace{-12pt}
\section{Related Work} In [1], the goal of the suggested research was to create an intelligent model that may aid in the security of the IoT framework. In ~\cite{caviglione2020tight}, The authors conducted a review of current cybersecurity research and discovered two significant trends. First is the adoption of old, established approaches that are still in use in many applications. Another is Machine learning (ML) based techniques, which are becoming increasingly popular.
It also mentions that ML is now being utilised in both malware and security. In~\cite{keserwani2021smart}, the author proposed a solution of traditional intrusion detection which gathers and uses huge data containing irrelevant information, leads to increase the detection time and lowers the accuracy. The solution uses combination of Particle Swarm Optimization (PSO) and Grey Wolf Optimization (GWO). Features obtained were given to Random Forest (RF) for classification, where these models were able to achieve 99.66\% accuracy for multiclass classification. In~\cite{ullah2021design}, author showed ML techniques are not able to detect attack for unpredictable network technologies but Deep Learning (DL) model was able to solve this problem. DL models used was Convolution Neural Networks (CNN) along with Transfer Learning, which was used for multiclass and binary classification on the Bot-IoT dataset. In~\cite{alkahtani2021intrusion}, the author proposed a novel framework using IoTID20 dataset. Proposed framework was using a CNN, LSTM and a hybrid CNN-LSTM model. The dimensionality of the network dataset was lowered, and the PSO approach was used to optimise the suggested system. These models were able to obtain 96.60\%, 99.82\% and  98.80\%  accuracy respectively. In~\cite{fan2020iotdefender}, Fan et. al. proposed an alternate model as a single detection model for heterogeneous data using Federated Transfer Learning. The model provided improvement detection of unknown attacks with 91.93\% of detection accuracy as compared to traditional methods. The model also generated less False Positive rate than single centralized model.\\
Several improvements have been made to enhance the accuracy for intrusion detection for IoT.
However, simple  DL-based IDSs face several inherent challenges during attacks classification in an intrusion detection system for IoT. 
Dataset with high number of features and high amount of data increases the time required to train on a DL model. Some features can negatively impact the training process of the model which reduces the model's attack detection capability.
Table 1 shows the summary of most recent researches with special emphasis on use of feature selection and data imbalance issues. The proposed approach involves uses feature selection techniques and also handles the data imbalances. 
\begin{table*}[t!]
\caption{Different intrusion detection approaches\label{tab:table1}}
\setlength{\arrayrulewidth}{0.4mm}
\setlength{\tabcolsep}{1pt}
\renewcommand{\arraystretch}{1}
\begin{tabular}{|p{1 cm}|p{2.5cm}|p{3cm}|p{2cm}|p{3cm}|p{3.7cm}|p{1.8cm}| }
\hline
\textbf{S.No.} & \textbf{References} &  \textbf{Handling data imbalance} & \textbf{Feature Selection} & \textbf{Classification Technique} & \textbf{Attack model for Prediction} & \textbf{Dataset} \\
\hline
1 &  Roddaet. al.  ~\cite{rodda2018network} & Not used & Not used & Multi layer Perceptron & N/w Intrusion Detection System & NSL –KDD \\
\hline
2 & Yanmiao et. al. ~\cite{li2020robust}  & Not used & Not used & Multi-CNN fusion model & NIDS & NSL –KDD \\
\hline
3 &  Mohamed et. al. ~\cite{ferrag2020rdtids} & Three RDTIDS mode & Not used & Rules and Decision Tree & NIDS & Bot-IoT \\
\hline
4 & Torres et al. ~\cite{torres2016analysis} & Not used & Not used & RNN & NIDS  & NSL –KDD \\
\hline
5 & Keserwani et. al. ~\cite{keserwani2021optimal} & Not used & GWO –CSA & DSAE & NIDS & UNSW NB-15 \\
 \hline
6 & Tama et al. ~\cite{tama2016classifier} & Not used & PSO+ ACO+ GA & Rotation forest & Binary classification & UNSW NB-15 \\
  \hline
7 & Alkahtani et al. ~\cite{alkahtani2021intrusion} & Not used & PSO & LSTM & Binary classification & IoTID 20 \\
\hline
\end{tabular}
\end{table*}

\section{ Overview of Deep Learning and Background Methodologies}
 \par
Network-based techniques aim to examine network traffic patterns of infected devices. The use of network flow data to detect traffic from malware-infected devices has been the subject of this research and two DL models namely ANN and LSTM~\cite{benkhelifa2018critical} shall be used as base for building four models.
\vspace{-0.3cm}
\subsection{ANN Architecture}
ANN is a DL based algorithm with an input layer with n nodes, where n denotes the number of independent variables, is followed by m linked hidden layers with activation function nodes and an output layer with activation function nodes in the basic architecture~\cite{schmidhuber2015deep}. ANN uses adam optimization to estimate the error gradient from observations in the training dataset. Back propagation is then used to modify the model weights and bias estimates associated with the nodes in order to arrive at an optimal solution. 

\subsubsection{No. of hidden layers} Because there are so many hidden layers, nodes and connections, there are exponentially huge numbers of trainable weights (relative to n) that can be changed to learn non-linear patterns in the dataset. Additional weights can be added by adding extra layers and nodes. Models with a few layers and nodes are unable to extract information from highly non-linear patterns, which may lead to  over-fitting. There are certain heuristics for figuring out the no's of hidden layers present, but still the processing capacity limits the amount of hidden layers and nodes which may be chosen.

\subsubsection{Activation functions}  Activation functions compute the weighted total of inputs and biases to decide whether a neuron can fire or not. The technique translates the output signal from the preceding layer's inputs and weights into a format that may be utilised as input nodes in the hidden layer.

\subsubsection{Sigmoid Function} Because it delivers a probability based conclusion, the sigmoid function is a preferred activation function in the output layer for binary classification. The vanishing gradient problem, on the other hand, causes the sigmoid function to fail in correctly converging after repeated backward iterations propagation. The sigmoid activation function equation is shown in eq. 1.

         \begin{equation}
\label{deqn_ex1}
S(x) =  \frac {1}{1+e^{-x}}
\end{equation}
where e\textsuperscript{-x} is Euler's number.
\subsubsection{Relu activation Function} The Rectified linear unit (ReLu) function is regarded as a safe default activation function. ReLU is computationally light, allowing for quick neural network(NN) training and performed effectively in many NN applications. The equation is given below:

       \begin{equation}
\label{deqn_ex1}
ReLu = f(x) =  max(0,x)
\end{equation}

\subsubsection{Loss Function}  It's a way of determining how effectively the algorithm models the data. The loss function will provide a greater value if the forecasts are completely wrong. For classification scenarios involving two classes, the binary cross entropy loss function is the most widely used loss function. 

\subsubsection{Optimizer} By iteratively updating weights and bias estimation, back propagation is a technique employed by ANN to lower the output of the loss function.

\subsubsection{Batch size and epochs} Batch size is a hyper-parameter that refers to the amount of data sent through the model before the weights are changed. The number of epochs refers to the number of iterations that the model must go through in order to include the whole training dataset. To obtain convergence, neural nets usually require numerous iterations across the dataset. The number of epochs and batch size affect the algorithm's ability to locate the best answer. The dataset condition determines the batch size to be used. As a consequence, model convergence depends on adjusting the number of observations in a batch and the number of epochs.
\vspace{-1cm}
\subsection{LSTM Architecture}
Long short-term memory is a sort of Recurrent Neural Network(RNN), that makes recalling past knowledge easier. Here, LSTM overcomes the RNN's vanishing gradient problem. The LSTM algorithm is highly suitable for discovering, analysing and forecasting time series with uncertain lengths . The model is trained via back-propagation. The Components of LSTM are discussed below ~\cite{greff2016lstm, https://doi.org/10.48550/arxiv.2207.00424}.
\hspace{1cm}

Input gate: With an input gate, it determines which value from the input should be utilised to change the memory. The sigmoid function specifies which numbers are allowed to pass between 0 and 1, while the tanh function gives the data input weight, indicating its relevance on a scale of -1 to 1.
\hspace{1cm}

Forget Gate: This Gate defines which block information should be deleted. The sigmoid function determines this. It creates a number between 0 and 1 for each number in the cell state. It looks at the previous state and the content input for each number in the cell state.
\hspace{1cm}

Output gate: The input and memory of the block defines the output gate. The sigmoid function determines which values are permitted to pass through numbers between 0 and 1. The tanh function multiplies Sigmoid's output by the weightage provided to input values, establishing their relevance level, which range from -1 to 1. LSTMs are subject to the same hyper-parameters as ANNs.
\section{Proposed Methodology and Experimental Setup}
In this study, four two-staged ML algorithms\cite{sugi2020investigation} have been proposed for intrusion detection in IoT networks.
Existing methods include DL systems, however they either process all data, even if only a portion of it was involved in the attack. However, in this paper, the correlation and mutual information based feature selection models are used as in ~\cite{koch2018mutual}. These feature selection methods have the ability to select the features that are relevant. Accuracy is not affected if only the selected features are used to predict the result. Filter-based, wrapper-based and embedded-based feature selections were the three types of feature selection. The filter based has been chosen in this research because it is computationally light, which fulfills our need for making overall model lightweight~\cite{8675917}. Workflow for the proposed methodology is given in Fig. 2. The steps taken for building models are explained below.
\vspace{-2pt}
\begin{figure}[!t]
\centering
\includegraphics[width=3.5 in]{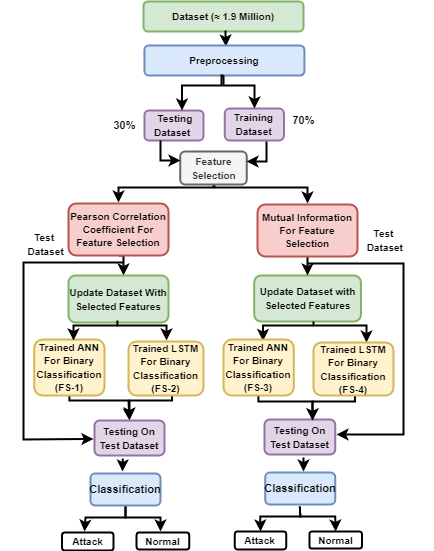}
\hfil
\caption{Proposed NFDLM workflow diagram}
\end{figure}
\vspace{-0.5cm}
\subsection{Dataset Description}
The BoT-IoT dataset~\cite{koroniotis2019towards} was developed at the Canberra Cyber Range Lab of the University of New South Wales. The data is separated into four botnet-based assaults and one 'benign' type. Majority of samples were contributed by DDos and DoS and due to that reason this dataset has been chosen to detect DDoS attack. Attack distribution for BoT-IoT dataset is given in Fig. 3. In this paper only two categories i.e. DDoS and benign are considered due to the negligible presence  of other attack categories. Therefore, there is an imbalance in the two categories. The dataset contains 42 features like bytes which represents total number of bytes in transactions, destination IP address and destination port numbers. The “spkts" implies which packet is from the source to the destination count to name few. The dataset is produced using five IoT devices. These devices are as follows: 
\begin{itemize}
  \item A weather station for measuring air pressure, humidity and temperature.
  \item Smart fridge used to adjust temperature below certain limit.
  \item Motion activated lights which switches on or off on the basis of randomly produced signal.
  \item Smart thermostat used to control building's temperature using air conditioner.
  \item Remotely activated  garage door: using probabilistic input opens or closes the door.
\end{itemize}
Dataset has 3.6 million samples which includes all the network traffic produced by these devices. In this paper, the selected DDoS attack type and benign type are selected from the sample which contains 1,726,624 samples for DDoS attack type and as dataset looks imbalanced. The number of  benign samples were only 477.
\begin{figure}[h!]
\centering
\includegraphics[width=3.3in]{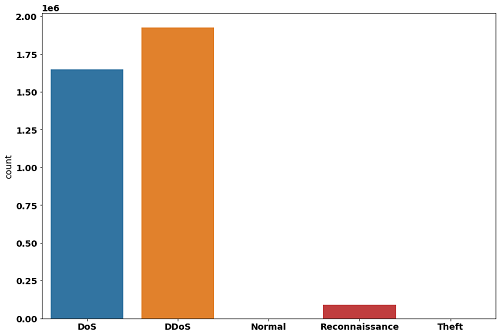}
\caption{Bar graph showing distribution of attack category}
\label{fig_1}
\end{figure}
\vspace{-0.5cm}
\subsection{Data Preprocessing}	In this stage, the dataset is getting preprocessed so that it  can be given to the classifier. Firstly all the features of type string object were dropped as their numerical representation in features are also present. There were features like pkSeqID which tells that row numbers so it was dropped as it was not relevant.“stime" and “ltime" tells record start time and end time so these were also dropped. After dropping 30 features were remaining. A copy of this updated dataset was saved. The Bot IoT dataset is unbalanced i.e. there were only 477 samples of benign type in comparison of 19,26,624 samples of DDoS attack type. To deal with data imbalance, "smote" was used which is minority over-sampling technique. After resampling we had 19,26,608 samples each for benign and attack categories. %
\subsubsection{Data Cleaning}
 In this research,the Python programming language is used to clean up the data. The categorical variables must be written as numeric values using API Scikit-learn and TensorFlow Keras API. One-hot encoding is used to transform the category variables in terms of numerical values if any. 
\subsubsection{Scaling and normalizing data}
In neural network software, scaling and normalising numeric input characteristics is widely employed to facilitate faster convergence and better outcomes. The Scikit-Learn API has four often used methods: min-max scaler, robust scaler, standard scaler and normalizer. Standard scaler method is used here. To discuss these methods the terms “x" and “xi"  are used to represent numerical feature vectors and  specific values inside the vectors respectively. The standard scaler is used to turn a feature vector into an approximation of a normal distribution with mean zero and unit variance. The standardization has been used to all the models.
         
         \begin{equation}
\label{deqn_ex1}
StandardScaler(xi, x) =  \frac {xi - mean(x)}{{  stdev(x) }}
\end{equation}

\subsubsection{Imbalances in Data Labels}
The response variables in the Bot-IoT dataset were unbalanced causes the data-imbalances which are a typical occurrence that can be troublesome. There are few, if any malicious flows in network traffic prior to botnet attack then after infection, majority of flow becomes infected. There are a variety of ways that may be utilised to vanish the negative effects of ML classifiers built on data sets which are imbalanced. Some of these tactics include up-sampling minority response variables or down-sampling majority response variables, the use of class weights and synthetic data creation. Synthetic Minority Over-sampling is employed in this model (SMOTE)~\cite{chawla2002smote}. it is a procedure in which SMOTE creates more minority class observations using the K-nearest neighbours method. This method has been proved beneficial in various machine learning case studies. SMOTE algorithms are included in the Scikit-Learn API.
\subsection{Feature Selection and Model Building} For each form of assaults, it chooses the most significant features. By lowering the dimensional of the feature, this module is necessary for the creation of a lightweight detection system with outstanding detection performance.
\par Feature Selection is applied to reduce the input dimension ~\cite{7762123} which in turn helps our model to converge faster as sometimes some features can have negative impact which leads to increases time to reduce loss function.

\subsubsection{Correlation} A variable's correlation indicates how that variable is related to the target. If the correlation between a variable and the target is “1", It indicates they're highly connected and constrained by direct proportionality. If  “-1", it implies they are inversely linked. For this model, Pearson correlation coefficient (equation 2) is used which checks the correlation between the two variables.

\begin{equation}
r = \frac {\sum\left(x_{i}-\bar{x}\right)\left(y_{i}-\bar{y}\right)} {\sqrt{\sum\left(x_{i}-\bar{x}\right)^{2} \sum\left(y_{i}-\bar{y}\right)^{2}}}
\end{equation}
\\ r =  \text{correlation coefficient}
\\ $x_{i}$=  \text{values of the x-variable in a sample} 
\\$\bar{x}$=  \text{mean of the values of the x-variable}
\\ $y_{i}$=  \text{values of the y-variable in a sample}
\\$\bar{y}$=  \text{mean of the values of the y-variable}
\subsubsection{Mutual information} Mutual information is a metric that assesses how much information can be extracted from one random variable given another. It assesses how much knowledge of one variable decreases uncertainty about another ~\cite{peng2005feature}.
It's a two-step process: Firstly, It ranks features according to a set of criteria/metrics and then the next step is to Choose the attributes that have the highest rating.
 
\par Table 2 shows the naming convention opted for our models. using “Feature selection 1"(FS1 model) which was using correlation coefficient as feature selection and ANN model as classifier. “Feature selection 2"(FS2 model) which was using mutual information as feature selection and ANN model as classifier. “Feature Selection 3"(FS3 model) which was using correlation coefficient as feature selection and LSTM model as classifier. “Feature selection 4"(FS4 model)~\cite{soe2020machine} which was using mutual information as feature selection and LSTM model as class predictor. These two-feature selection approaches will be utilised to select the appropriate feature set for each sort of attack, such as FS-1, FS-2, and so on. After obtaining the most relevant feature sets, the Deep learning (DL) algorithms in the “Model Trainer" module will generate the trained model for each type of attack. ANN and LSTM will be employed to predict the class label.

\vspace{-0.35cm}
\subsection{Training Model}
The Model Selector: This module develops and assesses a variety of ML methods before selecting the optimal one based on detection accuracy and other evaluation parameter for each of the sub-engines. Parameter tuning is done for the all models and best four are selected for attack detection and performance evaluation. Firstly, a base ANN model is built without any feature selection for comparative purpose. This model has input layer with 6 neurons and ReLu as an activation function, one dense layer with 6 neurons and ReLu as the activation function, and an output layer with Sigmoid as the activation function. Adam optimizer was used with binary cross-entropy as loss function. The model was trained with a batch size of 20 epochs and a batch size of 10 epochs.
\par For FS1 model, all the parameter were same as base ANN model it just had addition of correlation coefficient feature selection to reduce the no. of features which reduced to 9 feature only. With a batch size of 20 and 20 epochs, this model was trained.
\par For FS2 model, all the parameter were same as base ANN model it just had addition of Mutual Information feature selection to reduce the no of features reduced to 11 features. This model was trained using 20 epochs and a batch size of 20.
\par For FS3 model, all prepossessing were same as FS1 and FS2 model, feature selection was able to reduce the features to 13 feature.
Classic LSTM was modeled to maintain overall architecture lightweight. The model had three layers, the input layer had 64 neurons with ReLU as the activation function, while the second layer had 128 neurons with ReLU as the activation function, the output layer was a dense layer with one neuron and a Sigmoid activation function. The model was trained with 50 epochs and batch size of 32 and the Loss function chosen was binary cross-entropy with Adam optimizer.
\par For FS4 model all prepossessing  were  same as FS3 model. Here, mutual information was used as feature selector which reduces no. of features to 11. Classic LSTM was implemented to maintain overall architecture lightweight. The model had three layers. The input layer had 64 neurons with ReLU as the activation function, while the second layer had 128 neurons with ReLU as the activation function. The output layer was a dense layer with one neuron and a sigmoid activation function. model was trained with 50 epochs and batch size of 32. Loss function chosen was binary cross-entropy with Adam optimizer.

\vspace{-.2cm}
\section{Results/Discussion}
In this experiment, feature selector algorithms were able to reduce the number of features drastically. Feature selector for the FS1 selected 20 features which were highly correlating with other features so all of them were dropped. The threshold of 65\% had selected so that the feature correlating with other features with more than 65\% were dropped. The per protocol average rate per destination IP feature was substantially correlated with the “Rate" feature, which indicates a transaction's total number of packets per second. Because the feature “maximum length" of aggregated data is significantly associated with the mean duration of aggregated records, the “mean" feature was omitted. Similarly, it was find that another 18 features were closely correlated with each other as it can be seen by heat diagram shown in Fig. 4, all of the highly correlated features were removed. The feature selection in both FS1 and FS3 was the same as both were applied on the same data set, therefore, the features dropped in both were the same. A heat diagram is given in Fig. 4 which shows the features correlation with other features, darker the shade means the features are highly correlated.
\begin{table*}[ht!]
\caption{Models comparison\label{tab:table2}}
\begin{center}
\setlength{\arrayrulewidth}{0.4mm}
\setlength{\tabcolsep}{2pt}
\renewcommand{\arraystretch}{1}
\begin{tabular}{ |p{2cm}|p{2cm}|p{3cm}| p{3cm}|p{3cm}|p{4cm} | }
\hline
\textbf{Model} &  \textbf{Accuracy} & \textbf{Training Time}  & \textbf{Features} & \textbf{Classifier} & \textbf{Feature Selector}  \\
\hline
FS-1  & 98.53 & 1029 & 9 & ANN & Correlation Coefficient \\
FS-2 & 99.99   & 944 & 11 & ANN & Mutual Information \\
FS-3 & 99.89 & 2395 & 13 & LSTM & Correlation Coefficient \\
FS-4 & 99.84 & 2601 & 11 & LSTM & Mutual Information \\
\hline
\end{tabular}
\end{center}
\end{table*}
\vspace{-2pt}
\begin{table*}[ht!]
\caption{Effectiveness comparison
\label{tab:table3}}
\begin{tabular}{ |p{3cm}|p{4cm}|p{4cm}|p{3cm}|p{2cm}|}
\hline
\textbf{System} &  \textbf{Deep Learning Techniques}  & \textbf{Task} & \textbf{Dataset Used} & \textbf{Accuracy} \\
\hline
Jia et. al. ~\cite{jia2020flowguard} & LSTM & Binary Classification & CIC-DDoS2019 & 98.9 \\
\hline
Li et. al. ~\cite{li2020rtvd} &  LSTM & Binary Classification & CIC-DDoS2019 dataset &98.1\\
\hline

B. Roy et. al.~\cite{roy2018deep}&  Bi- LSTM & Binary Classification & UNSW-NB15 & 95.75%
\\
\hline
M. A. Ferrag et. al. ~\cite{ferrag2020deep}
&  RNN & Binary Classification & Bot-IoT dataset & 98.20
\\
\hline
NFDLM &  ANN + Mutual Information & Binary Classification  & Bot-IoT dataset &  99.99  
\\
\hline
\end{tabular}
\end{table*}

\begin{figure}[ht!]
\centering
\includegraphics[width=3.5 in]{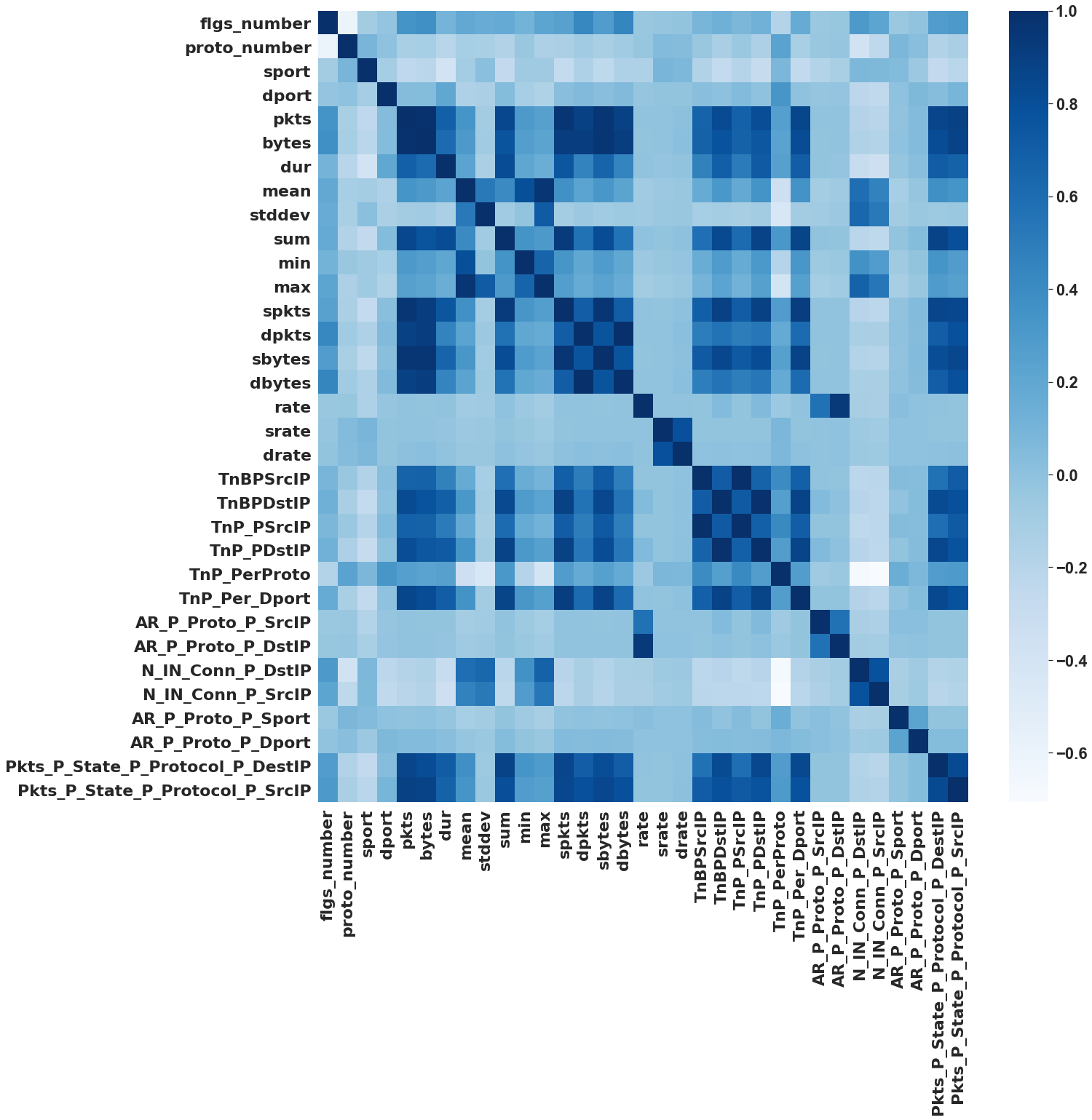}
\caption{Heat diagram for Correlation Coefficient feature selector}
\label{fig_1}
\end{figure}
\par The energies of all features for a feature selector are based on mutual Information and then the top 11 features are selected which are contributing in highest to classifying the attack type. First place has gone to the \textit{number of inbound connections per destination IP}, indicating that it had the most important role in attack identification. \textit{Total number of packets per protocol} feature was ranked second, while \textit{total number of bytes per destination IP} was ranked third. There is a graph provided in Fig. 5 which shows ranking of feature from the highest to the lowest. Both FS2 and FS4 were using mutual information as feature selector on the same data set hence same features were selected. These Features were used in classifier to detect attack. 
\vspace{-2pt}
\begin{figure}[ht!]
\centering
\includegraphics[width=3.5 in]{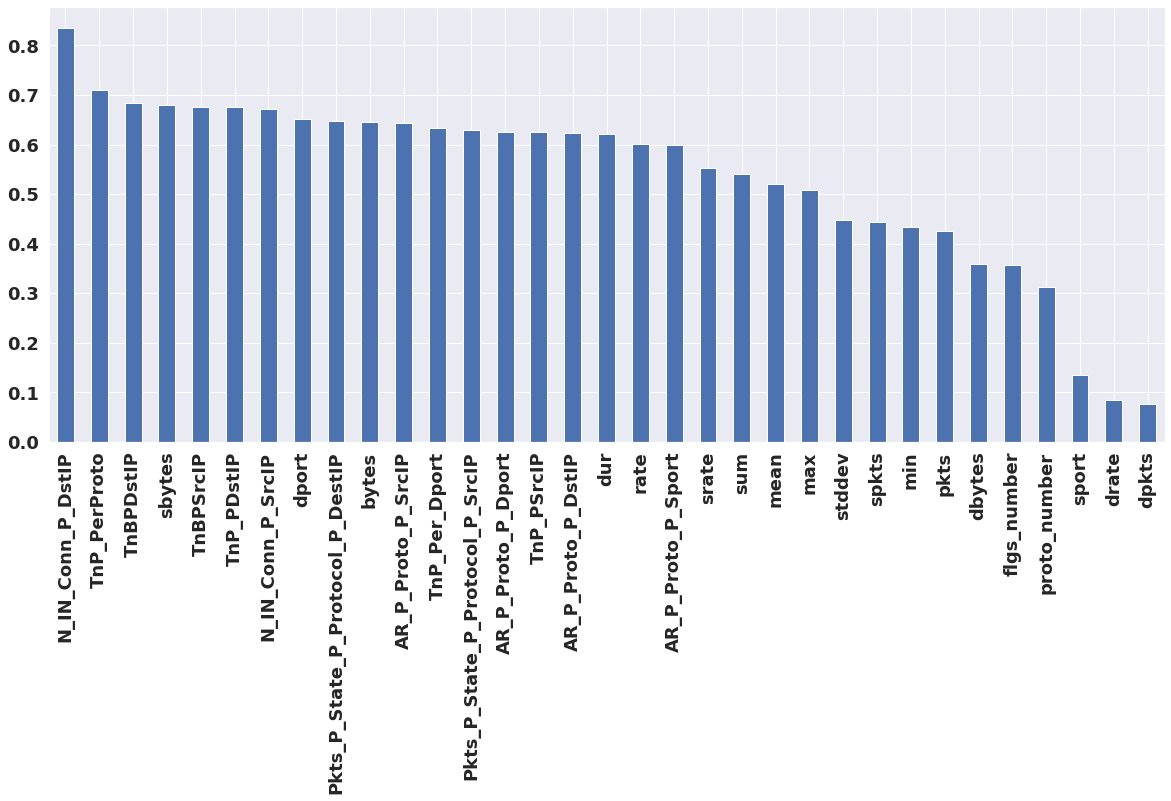}
\caption{Graph showing ranking of features based on Mutual Information}
\label{fig_1}
\end{figure}

\par For FS1 model (ANN + correlation coefficient) the accuracy obtained was 98.53 percent (table 2) and the Overall training time was 1029 seconds.

\par For FS2 model (ANN + Mutual Information),the accuracy obtained in this model was 99.99 percent (table 2) which was the best of our all the four models. Overall training time took 944 seconds which was the best among the four models and average epoch time achieved was 47.2 seconds.
\par For FS3 model (LSTM + Correlation Coefficient) the total training time was 2395 seconds. On an average, the epochs took 47.9 second which is better than FS-1 but lesser than FS2. A decent accuracy of 99.89 percent was achieved as shown in table 2.
\par For FS4 model (LSTM + Mutual Information) the accuracy was 99.84 percent but this model took the highest time to train. The total time to train was 2601 seconds with on average each epoch taking 52.02 seconds as shown in table 2.
\par Overall, it was discover that ANN with Mutual Information performed significantly better than the other models on the basis of accuracy and training time. FS-1 model performed well in terms of feature reduced. Features were reduced to 9 but accuracy achieved was poorest of all the models. FS-3 was the second best model in terms of accuracy. FS-4 performed decent in term of accuracy but it was taking the highest time to train. The Correlation Coefficients feature selector was fast in finding optimal feature in comparison with  Mutual Information as feature selector. Accuracy obtained by model with Mutual information feature selector was comparatively higher signifying  it's ability to select features mainly contributing in identifying DDoS attack. our models FS-2, FS-3, FS-4 were given superior result than recent studies both for DDoS i.e., binary classification (table 3). 
\vspace{-.25cm}
\section{Conclusion and Future Work}
In this research, four deep learning models are implemented for DDoS attack detection. We maintained lightweightness of NFDLM using feature reduction, selection and the implementation of basic ANN and LSTM. The FS2 model was chosen as it was performing better in both classification accuracy and time to train the model. In future, other deep learning models may be integrated to come up with a hybrid model to design a lightweight and hybrid NIDS with improved accuracy and training time. This experiment was not performed on industrial scale or real world IoT applications~\cite{Saurabh2020-je}, which is left as future work.
\vspace{-.2cm}
\bibliographystyle{IEEEtran}
\bibliography{bibliography.bib}

\newpage

\vfill

\end{document}